\documentstyle[12pt]{article}
\title{\baselineskip=9mm
Effects of finite width of excited states on heavy-ion sub-barrier fusion 
reactions}
\author{K. Hagino,$^{1,2}$ and N. Takigawa,$^{2}$
\\ \\
\medskip
{\it $^1$ Institute for Nuclear Theory, University of Washington, }\\
{\it Seattle, Washington 98915 }\\
{\it $^2$ Department of Physics,
Tohoku University, Sendai 980--8578, Japan}\\
}
\date{}

\begin{document}
\maketitle

\begin{center}
{\bf Abstract}
\end{center}

We discuss the effects of coupling of the relative motion to 
nuclear collective excitations which have a finite lifetime 
on heavy-ion fusion reactions at energies near and below the 
Coulomb barrier. 
Both spreading and escape widths are explicitly taken into 
account in the exit doorway model. 
The coupled-channels equations are numerically solved to 
show that the finite resonance width always hinders fusion cross 
sections at subbarrier energies irrespective of the relative 
importance between the spreading and the escape widths. 
We also show that the structure of fusion barrier distribution 
is smeared due to the 
spreading of the strength of the doorway state. 

\medskip

\noindent
PACS number(s):
25.70.Jj, 24.10.Eq,24.30.Cz,25.70.Mn

\newpage

\begin{center}
{\bf I. INTRODUCTION}
\end{center}

Extensive experimental as well as theoretical studies during the past 
two decades have led to a well-established idea that cross sections 
for heavy-ion fusion reactions are considerably enhanced at sub-barrier 
energies compared with predictions of the one dimensional potential 
model \cite{B88}. It has subsequently been concluded that 
these enhancements of fusion cross sections can be attributed to 
couplings between the relative motion 
of the colliding nuclei and several nuclear collective motions as well 
as transfer reactions \cite{BT98}. 
A standard way to address the effects of channel coupling on fusion 
cross sections is to numerically solve the coupled-channels equations. 
In coupled-channels calculations involving 
low-lying collective excitations of medium mass nuclei, 
which show the large enhancements of fusion cross sections, 
the excited states are usually assumed to have an infinite lifetime. 
However, when the excitation energies of the collective modes 
exceed the threshold energy for particle emission or 
the typical excitation energy of incoherent modes of excitations, 
like giant resonances in stable nuclei, they have 
finite lifetimes \cite{BB94}. 
Although the excitation energy of giant resonances in stable 
nuclei is in general 
so large that the effects of their excitations and thus their width 
can be well described 
by a static potential renormalisation \cite{THAB94,HTDDL97}, 
the effects of the finite width of excited states are expected 
to become important in discussing fusion reactions 
of weakly bound nuclei like $^{6,7}$Li, $^{9}$Be 
\cite{SYK83,HS92,RGMT97,TMS97} or 
nuclei far from the stability lines\cite{YSF96,HPCD92,CDLH95,TKS93,DV94}

Some attempts to include the effects of finite width 
in coupled-channels calculations have been made recently by 
Hussein {\it et al.} \cite{HT94,HPT95}. 
They used the exit doorway model \cite{BM69,BT85} to discuss the 
effects of spreading width $\Gamma^{\downarrow}$ on fusion reactions. 
Instead of numerically solving the resultant coupled-channels 
equations as they are, 
they introduced the constant coupling approximation to diagonalise 
the coupling matrix\cite{DLW83} 
and claimed that the spreading width further enhances fusion cross sections 
compared with the case where the excited state has an infinite life time. 
As for the effects of escape width $\Gamma^{\uparrow}$, they 
took another model, i.e. a model which uses 
a dynamical polarization potential to account for 
the effects of the break-up reaction. 
They thus showed that the escape width strongly reduces fusion 
cross sections. 

Although the results of Refs. \cite{HT94,HPT95} are interesting, 
there are still some unsatisfactory aspects in their approach. 
First the constant coupling approximation used in Refs. \cite{HT94,HPT95} 
does not provide satisfactory results in heavy-ion fusion reactions 
where the coupling extends outside the Coulomb 
barrier\cite{DL87,DNA92,HTB97}. 
Furthermore, the constant coupling approximation leads to complex 
values of eigen-energies as well as weight factors if one eliminates 
the internal degrees of freedom which couple to the doorway state. 
Refs. \cite{HT94,HPT95}, however, did not fully take this fact 
into consideration. 
The exact coupled-channels calculations 
are, therefore, urged in order to draw a definite conclusion on the 
effects of spreading width $\Gamma^{\downarrow}$ on heavy-ion 
fusion reactions. 
Secondly, the effects of escape width $\Gamma^{\uparrow}$ 
are not transparent in the polarization potential formalism used 
in Refs. \cite{HT94,HPT95}. Also it will be hard to evaluate 
the polarization potential for each reaction. 
For example, in fusion reactions of $^9$Be, 
most of the states of $^9$Be which are 
excited during fuion 
will eventually decay into the $n + 2\alpha$ channel, 
because the separation energy 
of $\alpha$ particle is small in this nucleus. 
It is not so straight forward to derive 
a poralization potential in a reliable way 
for such four-body problems. 

In this paper, we extend the model in Refs. \cite{HT94,HPT95} 
to treat both spreading and escape widths on an equal footing. 
This enables us to see explicitly the effects of both 
widths on subbarrier fusion reactions and to easily discuss 
the case where both of them are present simultaneously. 
We then carry out exact coupled-channels calculations to investigate 
the effects of the finite width on heavy-ion fusion reactions. 
It will be shown that 
both spreading and escape widths reduce fusion cross sections, 
contrary to the conclusions in Refs. \cite{HT94,HPT95}. 
We also discuss the effects of the finite width on the fusion barrier 
distribution defined as the second derivative of the product of 
the fusion cross section 
and the center of mass energy, $E \sigma$ \cite{RSS91,LDH95}. 

The paper is organised as follows. In Sec. II, we briefly review 
the exit doorway model and derive coupled-channels equations which 
account for the effects of the finite width of excited states. We 
consider three cases, i) where only the spreading width 
exists, ii) only the escape width exists, and iii) both 
the spreading and escape widths are present simultaneously. 
In Sec. III, we present numerical results of the 
coupled-channels calculations for these three cases and discuss the 
effects of finite width on fusion cross sections as well as on fusion 
barrier distributions. The summary is given in Sec. IV. 
Finally in Appendix A, a time dependent approach to discuss the 
effects of escape width is presented. 

\medskip

\begin{center}
{\bf II. COUPLING TO RESONANCE CHANNEL}
\end{center}

\begin{center}
{\bf A. Effect of spreading width}
\end{center}

We first discuss the effects of the spreading width on sub-barrier 
fusion reactions. To this end, we use the exit doorway 
model \cite{HT94,HPT95,BM69,BT85}. 
In this model, the relative motion between the colliding nuclei 
couples to many excited states only through the doorway state, 
i.e. the collective state. 
We assume the following 
Hamiltonian for the fusing system: 
\begin{equation}
H=
-\frac{\hbar^2}{2\mu}\frac{d^2}{dr^2}
+\frac{J(J+1)\hbar^2}{2\mu r^2}
+V_N(r)+\frac{Z_PZ_Te^2}{r}+H_{int}(\xi)+V_{coup}(r,\xi),
\end{equation}
where $r$ is the coordinate of the relative motion 
between the projectile and the target, and $\mu$ is the reduced mass.
$V_N$ is the bare nuclear potential, $Z_P$ and $Z_T$ are the 
atomic numbers of the projectile and the target, respectively. 
$H_{int}$ describes the intrinsic excitations in one of the 
colliding nuclei, 
and $V_{coup}$ the coupling between these excitations (generically 
denoted by $\xi$) and the relative motion. 
In writing Eq. (1), we used the no-Coriolis 
approximation and replaced the angular momentum operator for the 
relative motion by the total angular momentum $J$ \cite{TI86,HTBB95}. 
Following Refs. \cite{HT94,HPT95,BT85}, we assume that the 
intrinsic Hamiltonian $H_{int}$ and the coupling Hamiltonian 
$V_{coup}$ are given respectively by 
\begin{equation}
H_{int}= |d>E_d<d| + \sum_j|j>e_j<j|
+\sum_j\left[ |j> \Delta_j <d| + |d> \Delta^*_j <j| \right], 
\end{equation}
\begin{equation}
V_{coup}=f(r)\left(|0><d| + |d><0| \right),
\end{equation}
where $|0>, |d>$, and $|j>$ denote the ground state, the doorway 
state, and the other intrinsic states, respectively. 
$E_d$ and $e_j$ are the energy of the doorway state and that 
of the state $|j>$, respectively. 
In our example in this subsection where we discuss 
the effects of spreading width, $|j>$ are 
uncorrelated $1p1h$ states or more complicated many particle 
many hole states within the same nucleus. The 
former yields the Landau damping, while the latter the spreading 
width \cite{BB94}. 
In Eqs. (2) and (3), $\Delta$ and $f(r)$ are 
the coupling strength between the doorway state and $|j>$, 
and between the doorway state and the ground state, respectively. 
Inherently, the former is independent of $r$. 
For simplicitly we have assumed that the doorway state $|d>$ linearly 
couples to the ground state $|0>$. The extension 
to the case where there exist higher order 
couplings \cite{HTD97} is straight forward. 

The intrinsic Hamiltonian $H_{int}$ can be diagonalised by 
introducing the normal modes $|\varphi_i>$ by 
\begin{eqnarray}
H_{int}&=&\sum_i |\varphi_i> E_i <\varphi_i|, \\
|\varphi_i>&=&\alpha_i |d> + \sum_j \beta_{ij} |j>.
\end{eqnarray}
When the states $|j>$ are distributed with equal 
energy spacing from $-\infty$ to $\infty$, and the coupling 
strengths $\Delta_j$ are independent of $j$, i.e., 
\begin{eqnarray}
e_j&=&jD~~~~~(j=0,\pm 1, \pm 2, \cdots), \\
\Delta_j &=& \kappa,
\end{eqnarray}
$E_i$ and $\alpha_i$ in Eqs. (4) and (5) are given by 
\begin{eqnarray}
E_i&=& E_d + \frac{\pi\kappa^2}{D}\cot\frac{\pi E_i}{D}, \\
\left|\alpha_i\right|^2&=& \frac{1}{\frac{2\pi}{D}}
\frac{\Gamma^{\downarrow}}{(E_i-E_d)^2
+\frac{\Gamma^{\downarrow 2}}{4}},
\end{eqnarray}
respectively (the Breit-Wigner distribution) \cite{BM69}. 
Here $\Gamma^{\downarrow}$ is the spreading width, including the Landau 
width, and is defined by 
$\Gamma^{\downarrow}=2\pi\kappa^2/D$. In obtaining Eq. (9), we assumed 
that the coupling strength $\kappa$ is much larger than the energy 
spacing $D$ and neglected a term of the order of $D^2/\kappa^2$ 
in the denominator. 

Expanding the total wave function with the eigen-states $|\varphi_i>$ as 
\begin{equation}
\Psi(r,\xi)=\frac{u_0(r)}{r}|0> + \sum_i\frac{u_i(r)}{r}|\varphi_i>,
\end{equation}
the coupled-channels equations read 
\begin{eqnarray}
&&\left[
-\frac{\hbar^2}{2\mu}\frac{d^2}{dr^2}
+\frac{J(J+1)\hbar^2}{2\mu r^2}
+V_N(r)+\frac{Z_PZ_Te^2}{r}-E\right]u_0(r) 
+\sum_j\alpha_j f(r) u_j(r)=0, \\
&&\left[
-\frac{\hbar^2}{2\mu}\frac{d^2}{dr^2}
+\frac{J(J+1)\hbar^2}{2\mu r^2}
+V_N(r)+\frac{Z_PZ_Te^2}{r}-E+E_j\right]u_j(r) 
+\alpha^*_j f(r) u_0(r)=0. \nonumber \\
\end{eqnarray}
These equations may be solved by imposing the incoming wave 
boundary condition inside the Coulomb barrier, i.e. 
\begin{eqnarray}
u_i(r)&=& T_i \exp\left(-i \int^r_{r_{abs}} k_i(r')dr'\right)
~~~~~~~~~~~r\leq r_{abs}, \\
&=&H_J^{(-)}(k_ir)\delta_{i,0} + R_i H_J^{(+)}(k_ir)
~~~~~~~~r\to\infty,
\end{eqnarray}
where $k_i(r)$ is the local wave length for the $i$-th channel, and 
$r_{abs}$ is the absorption radius where the incoming wave boundary 
condition is imposed. 
$H_J^{(-)}$ and $H_J^{(+)}$ are the incoming and the outgoing 
Coulomb waves, respectively. The fusion cross section is then 
obtained as 
\begin{equation}
\sigma(E)=\frac{\pi}{k^2}\sum_J(2J+1)
\left(
\frac{k_0(r_{abs})}{k}
|T_0|^2 + \sum_i \frac{k_i(r_{abs})}{k}|T_i|^2\right),
\end{equation}
where $k=k_0(\infty)$ is the wave length of the entrance channel 
at $r\to\infty$. 
Note that in  contrast to the works by Hussein {\it et al.} which 
eliminate the intrinsic degree of freedom $|j>$ 
and introduce the complex $Q$-value \cite{HT94,HPT95}, we treat 
it explicitly in the coupled-channels calculations. 

\begin{center}
{\bf B. Effect of escape width}
\end{center}

We next consider the effects of escape width. In this case, the 
states $|j>$ represent particle continuum states. 
The coupled-channels equations to be solved are exactly the same as 
Eqs. (11) and (12). 
One may imagine the box normalisation of these 
states in order to match with 
the discretisation of the continuum states $|j>$. 
The difference appears at the final stage. We define the 
complete fusion as such a process, where the whole projectile is 
absorbed by the whole target without emitting any particle 
prior to the fusion. 
Therefore, the particle continuum states $|j>$ have to be excluded 
from the final states in obtaining the cross section for the complete 
fusion. 
Since the probability to find a state $|j>$ in the state for the $i$-th 
normal mode $|\varphi_i>$ is given by $|\beta_{ij}|^2$, 
the complete fusion cross section is given by 
\begin{equation}
\sigma(E)=\frac{\pi}{k^2}\sum_J(2J+1)
\left(\frac{k_0(r_{abs})}{k}|T_0|^2 + \sum_i |\alpha_i|^2
\frac{k_i(r_{abs})}{k}|T_i|^2\right).
\end{equation}
In deriving Eq. (16), we used the normalisation condition 
of the normal states, i.e. $|\alpha_i|^2=1-\sum_j|\beta_{ij}|^2$. 

Another approach to fusion reactions in the presence of a 
break-up channel is to evaluate the loss of flux during 
fusion \cite{HPCD92,CDLH95,TKS93}. 
The relation between such approach and the present formalism 
is given in Appendix by using a time dependent theory. 

\begin{center}
{\bf C. Interplay between spreading and escape widths}
\end{center}

Lastly we consider the case where both spreading and escape 
widths are present simultaneously and interplay with each other. 
In this case, the intrinsic Hamiltonian is given by 
\begin{eqnarray}
H_{int} &=& |d>E_d<d| 
+ \sum_j|j^{\uparrow}>e^{\uparrow}_j<j^{\uparrow}|
+ \sum_j|j^{\downarrow}>e^{\downarrow}_j<j^{\downarrow}| \nonumber \\
&&+\sum_j\left[ |j^{\uparrow}> \Delta^{\uparrow}_j <d| 
+ |d> \Delta^{\uparrow *}_j <j^{\uparrow}| \right]
+\sum_j\left[ |j^{\downarrow}> \Delta^{\downarrow}_j <d| 
+ |d> \Delta^{\downarrow *}_j <j^{\downarrow}| \right], \nonumber \\
\end{eqnarray}
where $|j^{\uparrow}>$ and $|j^{\downarrow}>$ 
represent particle continuum states and complicated particle hole 
bound states, respectively. 
We diagonalise this Hamiltonian by introducing the normal 
states defined by 
\begin{equation}
|\varphi_i>=\alpha_i |d> + \sum_j \beta^{\uparrow}_{ij} |j^{\uparrow}>
+\sum_j \beta^{\downarrow}_{ij} |j^{\downarrow}>.
\end{equation}
As in Sec. II.A, if we assume a uniformely spaced sequence 
of energies $e_j^{\uparrow}$ and $e_j^{\downarrow}$ and state independent 
coupling strengths $\Delta_j^{\uparrow}$ and $\Delta_j^{\downarrow}$, 
the eigen-values $E_i$ and the doorway amplitude $\alpha_i$ are 
given by 
\begin{eqnarray}
E_i&=& E_d + \frac{\pi(\kappa^{\uparrow 2}+\kappa^{\downarrow 2})}{D}
\cot\frac{\pi E_i}{D}, \\
\left|\alpha_i\right|^2&=& \frac{1}{\frac{2\pi}{D}}
\frac{\Gamma^{\uparrow}+\Gamma^{\downarrow}}
{(E_i-E_d)^2+\frac{(\Gamma^{\uparrow}+\Gamma^{\downarrow})^2}{4}},
\end{eqnarray}
respectively. We have assumed that the energy spacing of 
the particle continuum states $|j^{\uparrow}>$ is the same as 
that of the particle-hole bound states $|j^{\downarrow}>$. 
Here $\Gamma^{\uparrow}$ and $\Gamma^{\downarrow}$ are 
escape and spreading widths given by $2\pi\kappa^{\uparrow 2}/D$ 
and $2\pi\kappa^{\downarrow 2}/D$, respectively. 
Excluding the particle continuum states $|j^{\uparrow}>$ from 
the final states, the complete fusion cross section is given by 
\begin{equation}
\sigma(E)=\frac{\pi}{k^2}\sum_J(2J+1)
\left[\frac{k_0(r_{abs})}{k}|T_0|^2 + \sum_i \left(|\alpha_i|^2+
\frac{\Gamma^{\downarrow}}{\Gamma^{\uparrow}+\Gamma^{\downarrow}}
\right)
\frac{k_i(r_{abs})}{k}|T_i|^2\right].
\end{equation}
In deriving Eq. (21), we used the identity 
$\sum_j|\beta_{ij}^{\downarrow}|^2=
\Gamma^{\downarrow}/(\Gamma^{\uparrow}+\Gamma^{\downarrow})$.
The complete fusion cross section is thus initimately related 
to the ratio 
$\Gamma^{\downarrow}/(\Gamma^{\uparrow}+\Gamma^{\downarrow})$. 

\medskip

\begin{center}
{\bf III. EFFECTS OF FINITE WIDTH ON FUSION CROSS SECTIONS AND 
BARRIER DISTRIBUTIONS}
\end{center}

We now present the results of our calculations of fusion 
cross sections and fusion barrier distributions. 
We consider fusion reactions between 
$^{16}$O and $^{144}$Sm in the presence of low-lying 
octupole phonon excitations in the latter nucleus. 
We artificially set its excitation energy $E_d$ to be 2 MeV and 
assume that it has a hypothetical total width of 1 MeV. 
Our conclusions do not depend so much on the particular 
choice for these parameters. 
We use the collective model for the coupling form 
factor $f(r)$, i.e. 
\begin{equation}
f(r)=\frac{\beta_3}{\sqrt{4\pi}}\left(-R_T\frac{dV_N}{dr} 
+\frac{3}{2\lambda+1}Z_PZ_T\frac{R_T^{\lambda}}{r^{\lambda+1}}
\right),
\end{equation}
where $\lambda=3$ is the multipolarity of the excitation and 
$R_T$ is the radius of $^{144}$Sm. $\beta_3$ is the deformation 
parameter of the phonon excitation, which was chosen to be 
0.205 with the target radius of $R_T=1.06 A^{1/3}$ fm. 
We used the same nuclear potential $V_N$ 
as that in Refs. \cite{HTDDL97,HTD97}, i.e. a Woods-Saxon 
potential whose depth, range parameter and surface 
diffuseness are $V$=105.1 MeV, $r_0$=1.1 fm, and $a$=0.75 fm, 
respectively. 
In the actual calculations, we introduced a cut-off energy and 
a finite energy spacing for the excited states by considering 
the normal states between $E_d - 1$ and $E_d + 1$ MeV with the 
energy spacing $D$ of 0.2 MeV. 
We have checked that our conclusions do not qualitatively alter 
when the cut-off energy is taken to be larger and/or the energy 
spacing smaller. 

We first discuss the effects of spreading width. 
The upper panel of Fig. 1 shows the fusion cross section of this 
system. The solid line was obtained by numerically solving 
the coupled-channels equations Eqs. (11) and (12) with 
$\Gamma^{\downarrow}$=1 MeV, while the dashed line is 
the result when the doorway state has an infinite lifetime. 
The figure also contains the result for the no coupling case 
(the dotted line) for comparison. 
One can see that the spreading width slightly reduces the fusion 
cross section, though it is still enhanced compared with 
the case of no coupling. Our result contradicts the 
conclusion in Refs. \cite{HT94,HPT95} where it was claimed that 
the spreading width enhances the fusion cross section. 
This discrepancy could be associated with 
the constant coupling model and/or 
the incorrect treatments of complex potentials and weight factors 
in Refs. \cite{HT94,HPT95}. 

The fact that the spreading width reduces 
the enhancement of fusion cross section 
can be understood in the following way. After eliminating the 
intrinsic states $|j>$, the coupled-channels problem 
given by Eqs. (11) and (12) reduces to the two channel problem 
with the coupling matrix \cite{HT94,HPT95} (see also Appendix)
\begin{equation}
\left(
\begin{array}{cc}
0&f(r)\\ f(r)&E_d-i\frac{\Gamma}{2}
\end{array}
\right).
\end{equation}
If we diagonalise this coupling matrix with a 
bi-orthogonal basis, the lower potential barrier 
is given by \cite{HT94,HPT95}
\begin{equation}
V_-(r)=V_N(r)+\frac{Z_PZ_Te^2}{r}
+\frac{1}{2}\left(E_d-i\frac{\Gamma^{\downarrow}}{2}
-\sqrt{E_d^2-\frac{\Gamma^{\downarrow 2}}{4}
+4f(r)^2-iE_d\Gamma}\right).
\end{equation}
The real part of the lower barrier thus always increases when the width 
$\Gamma$ is non-zero, leading to smaller penetrabilities at energies 
below the barrier. 
When the doorway energy $E_d$ is much larger than the coupling 
form factor $f(r)$, Eq. (24) is transformed to 
\begin{equation}
V_-(r)=V_N(r)+\frac{Z_PZ_Te^2}{r}
-f(r)^2\frac{E_d+i\frac{\Gamma^{\downarrow}}{2}}
{E_d^2+\frac{\Gamma^{\downarrow 2}}{4}},
\end{equation}
which is identical to the adiabatic barrier derived from Eq. (B.21) 
in Ref. \cite{BT85}. 

One might expect that fusion cross section is enhanced 
if the spreading width is finite, since the doorway state then couples 
to the environmental background whose energy is 
lower than that of the doorway state itself and thus 
the effective excitation energy of the doorway state becomes lower. 
However, this intuitive picture does not hold if there is non-negligible 
coupling $f(r)$ around the barrier position, which modifies the 
potential barrier according to Eq. (24). 
Since the fusion cross section is much more sensitive to the barrier 
height at energies well below the Coulomb barrier than 
the energy transfer which takes place before the projectile 
hits the Coulomb barrier, the net effects of the spreading width 
cause the hindrance of the fusion cross section. 
A similar importance of the potential renormalisation concerning the 
effects of transfer 
reaction with positive $Q$-value on subbarrier fusion reactions has 
been pointed out in Ref. \cite{L84}.

This contrasts to the situation where the doorway state couples to a small 
number of intrinsic states $|j>$. In such cases, 
the coupling could further enhance the fusion cross section, 
as is usually the case in double phonon couplings \cite{HKT98}. 
When the doorway state couples to a large number of surrounding states, as 
is discussed in this paper, the relaxation time becomes very short 
and consequently the couplings begin 
to reveal a dissipative character \cite{BT85}. 
The hindrance of fusion cross sections due to the spreading width 
thus resembles the dissipative quantum 
tunneling, which has been an extremely popular subject 
during the past two decades 
in many fields of physics and chemistry\cite{CL81,JJAP93}. 

The above conclusions have been reached by assuming the 
Breit-Wigner distribution given by Eq. (9) for the 
non-collective states $|j>$. 
In order to test the sensitivity to the property of the distribution, 
we repeated the calculations by assuming the Lorentzian distribution 
(not shown) and obtained the similar conclusions concerning the role 
of the finite resonance width in heavy-ion fusion reactions. 
The difference is negligible especially when the doorway energy $E_d$ 
is larger than the width $\Gamma$. 

The lower panel of Fig. 1 shows fusion barrier distributions for this 
system. This quantity is defined as the second derivative of $E\sigma$ 
with respect to the energy $E$ \cite{RSS91}, and has been experimentally 
shown to be very sensitive to the nuclear structure of the 
colliding nuclei\cite{LDH95}. 
We used the 3-point differece formula with 
an energy spacing of 1.8 MeV to obtain the second derivative 
from the fusion cross sections\cite{LDH95}. 
The meaning of each line is the same as that in the upper panel. 
When there exists no coupling between the ground and the doorway 
states, the barrier distribution has only a single peak, corresponding 
to a single potential barrier, i.e. the bare potential 
barrier. 
If the coupling is turned on, the single barrier splits to two and 
the barrier distribution has two peaks (the dashed line). 
This double peaked structure is somewhat smeared when the width of the 
doorway state is finite, since it re-distributes 
the strength of the doorway state (the solid line). 
As a consequence, the higher peak of the fusion barrier distribution 
becomes less apparent. 

We next discuss the effects of escape width. 
The solid line in Fig. 2 was obtained by setting 
$\Gamma^{\uparrow}$=1 MeV and $\Gamma^{\downarrow}$=0 MeV, 
and using Eq. (16). The dotted and the dashed lines 
are the same as in Fig. 1. 
As will be discussed in Appendix, the escape width is intimately 
related to a loss of flux due to the break-up reaction, and 
strongly hinders the fusion cross section over a wide range of 
bombarding energies. The fusion cross section is smaller 
even than that in the absense of the channel couplings at energies 
above the Coulomb barrier. The escape width also lowers the 
height of the main peak of the fusion barrier distribution and at 
the same time broadens the fusion barrier distribution. 

When there exist both spreading and escape widths 
simultaneously, one expects to have a situation intermediate 
between depicted in Figs. 1 and 2. 
In order to demonstrate this, Fig. 3 shows the result 
when the both widths are set to 0.5 MeV. 
Since both spreading and escape widths always reduce 
the fusion cross section, they are smaller 
than those in the infinite lifetime case over the entire 
energy range shown in Fig. 3. 
One finds that the degree of hindrance is intermediate 
between Figs. 1 and 2, as expected. 

\medskip

\begin{center}
{\bf IV. SUMMARY}
\end{center}

We have derived coupled-channels equations which take into 
account the effects of finite width of an excited state. 
This formalism treats the spreading and the escape widths 
on an equal footing, and thus enables one to discuss 
an interplay between them. 
To this end, we used the exit doorway model. 
Numerical solutions of the coupled-channels 
equations showed that both widths hinder the fusion 
cross section. The degree of hindrance is moderate 
when the spreading width dominates the total 
width, while the fusion cross section is considerably reduced 
in the opposite case, i.e. when the escape width dominates 
the total width. 
We also investigated the effects of finite width on the fusion 
barrier distribution. 
We demonstrated that the spreading width smears the structure of 
the fusion barrier distribution and also that the 
escape width lowers the height of the main peak of the fusion barrier 
distribution. 
These considerations are important when one analyses high precision 
measurements of the fusion reactions of fragile nuclei like 
$^{6,7}$Li or $^{9}$Be, which have been undertaken recently, or 
when one discusses fusion reactions of unstable nuclei. 
We will report analyses of these experimental data in a separate paper. 

\begin{center}
{\bf ACKNOWLEDGMENTS}
\end{center}

The authors thank M. Dasgupta, D.J. Hinde, and C.R. Morton for
useful discussions, and 
the Australian National University for its 
hospitality and for partial support for this project. 
The work of K.H. was supported by the Japan Society for the Promotion of
Science for Young Scientists.  
This work was also supported by the Grant-in-Aid for General                
Scientific Research,                                                        
Contract No.08640380, Monbusho International Scientific Research Program:   
Joint Research, Contract No. 09044051,                                      
from the Japanese Ministry of Education, Science and Culture. 

\bigskip

\begin{center}
{\bf APPENDIX A: 
TIME DEPENDENT APPROACH OF THE EFFECT OF ESCAPE WIDTH}
\end{center}

In this appendix, we discuss the relation between 
the coupled-channels formalism discussed in Sec. II.B and an 
approach using the flux loss during fusion 
due to a break-up reaction. 
To this end, we use a time dependent approach. 
Assuming that the total wave function at time $t$ is given by 
\begin{equation}
|\Psi (t)>=a_0(t)|0> + \sum_i a_i(t)e^{-iE_it/\hbar}|\varphi_i>,
\end{equation}
the time dependent coupled equations which corresponds to Eqs. 
(11) and (12) read \cite{CRHT94,BCHT96}
\begin{eqnarray}
\dot{a}_0(t)&=&\sum_i\frac{1}{i\hbar}\alpha_i f(r(t))e^{-iE_it/\hbar}
a_i(t), \\
\dot{a}_i(t)&=&\frac{1}{i\hbar}\alpha^*_i f(r(t))e^{iE_it/\hbar}
a_0(t). 
\end{eqnarray}
If we assume the Breit-Wigner distribution Eq. (9) for the 
incoherent states $|j>$, these two equations can be combined 
to give \cite{CRHT94,BCHT96}
\begin{equation}
\dot{a}_0(t)=-\frac{f(r(t))}{\hbar^2}\int^t_{-\infty}
dt'e^{-i(E_d-i\frac{\Gamma}{2})(t-t')/\hbar}a_0(t')
f(r(t')). 
\end{equation}
In deriving this equation, we have assumed that the energy spacing 
of the states $|j>$ is small enough and replaced the summation 
over the normal states $|\varphi_i>$ with an integration over the 
energy of these states $E_i$. 

Here we introduce the doorway amplitude by 
\begin{equation}
a_d(t)=\frac{1}{i\hbar}\int^t_{-\infty}
dt'e^{-i(E_d-i\frac{\Gamma}{2})(t-t')/\hbar}a_0(t')
f(r(t')). 
\end{equation}
Eq. (29) can then be written in a form of two coupled equations as 
\begin{eqnarray}
\dot{a}_d(t)&=&\frac{1}{i\hbar}\left(E_d-i\frac{\Gamma}{2}\right)a_d(t) 
+\frac{1}{i\hbar}f(r(t))a_0(t), \\
\dot{a}_0(t)&=&\frac{1}{i\hbar}a_d(t)f(r(t)).
\end{eqnarray}
These equations provide a two level problem 
with the coupling matrix given by Eq. (23) \cite{HT94,HPT95}. 

When the escape width dominates the total width, i.e. 
$\Gamma\sim\Gamma^{\uparrow}$, 
the survival probability of the system at time $t$ is given by 
\begin{equation}
P_s(t)=\left|<0|\Psi(t)>\right|^2 + \left|<d|\Psi(t)>\right|^2.
\end{equation}
Noticing that $<d|\Psi(t)>$ is nothing but the doorway 
amplitude $a_d(t)$, one can easily show that the time dependence 
of the survival probability is given by 
\begin{equation}
\frac{d}{dt}P_s(t)=-\frac{\Gamma^{\uparrow}}{\hbar}
\left|a_d(t)\right|^2.
\end{equation}
The survival probability $P_s(t)$ is thus a decreasing function of 
time $t$, and $1-P_s(t)$ represents the probability of the flux loss 
caused by the particle emission.

\newpage

\begin{center}
{\bf Figure Captions}
\end{center}

\noindent
{\bf Fig. 1:} 
Effects of the spreading width on the fusion cross section (the upper panel) 
and fusion barrier distribution (the lower panel). The dotted line 
is for the case of no coupling. The solid line takes 
into account the effects of coupling of the relative motion 
to a doorway state at 2 MeV with the width 1 MeV, while the dashed 
line assumes that the doorway state has an infinite lifetime. 

\medskip

\noindent
{\bf Fig. 2:} 
Same as fig.1, but for the escape width. 

\medskip

\noindent
{\bf Fig. 3:} 
Same as fig.1, but in the simultaneous presence of 
spreading and escape widths. Both widths are assumed to be 0.5 MeV. 

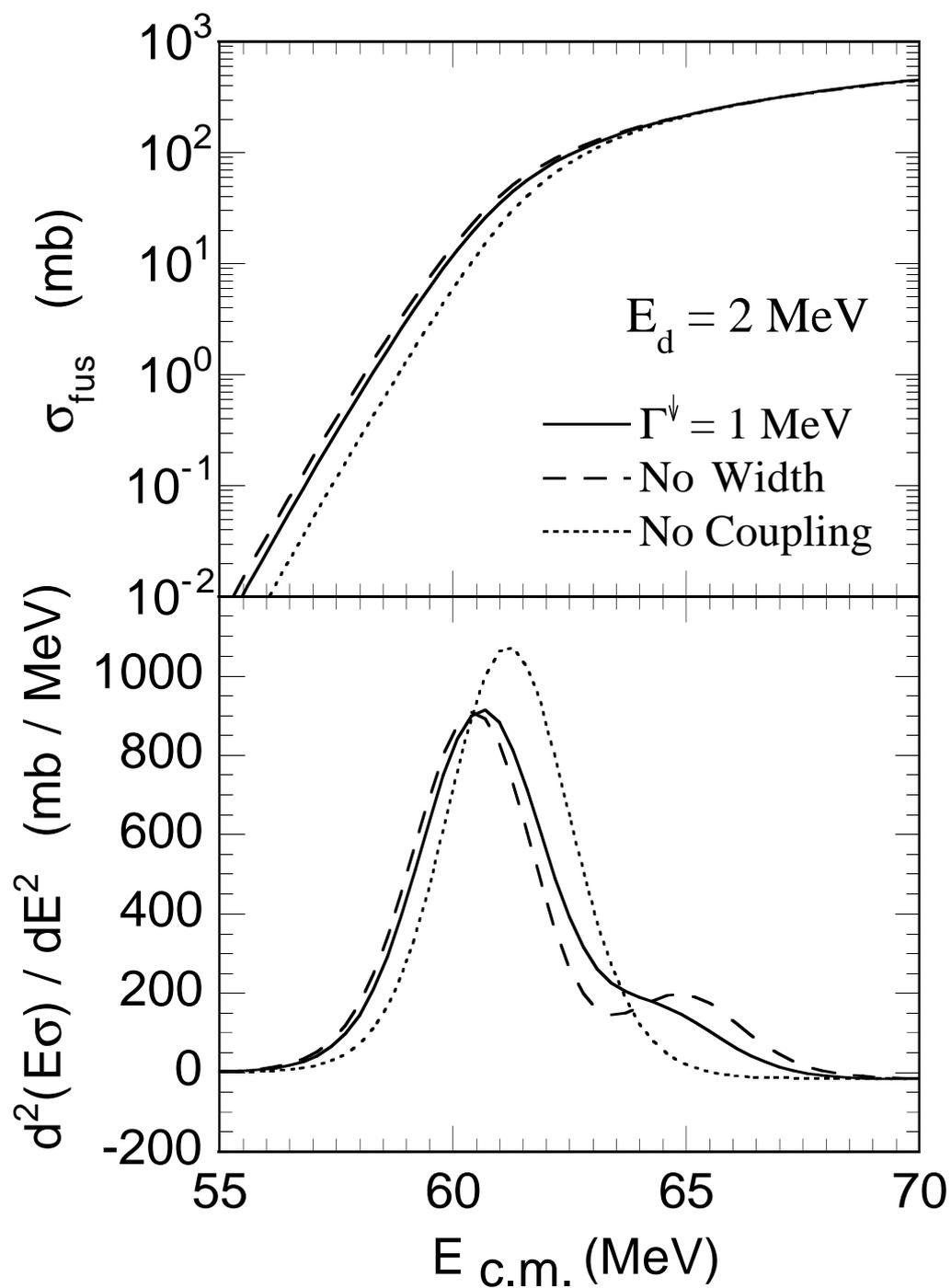
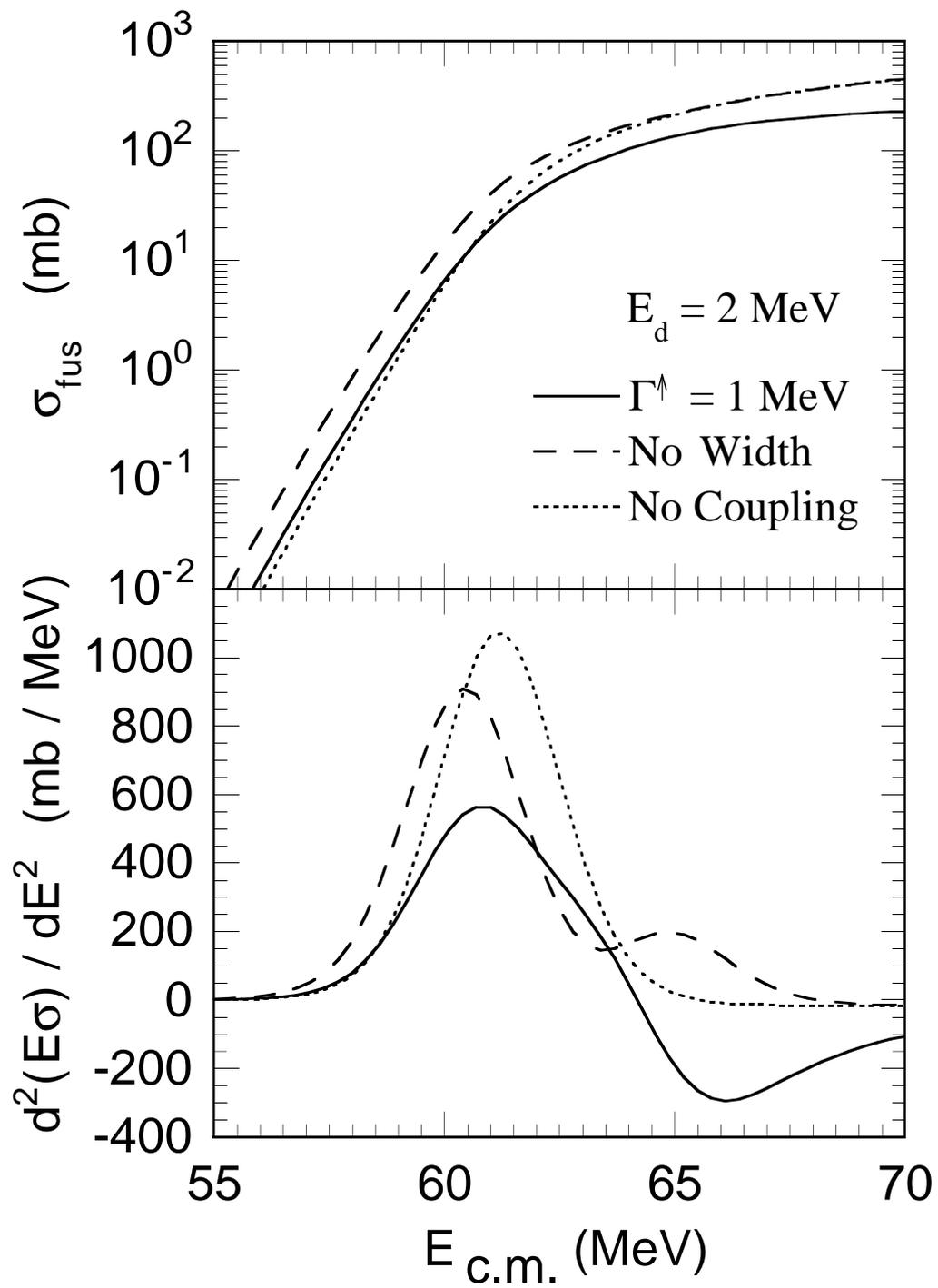
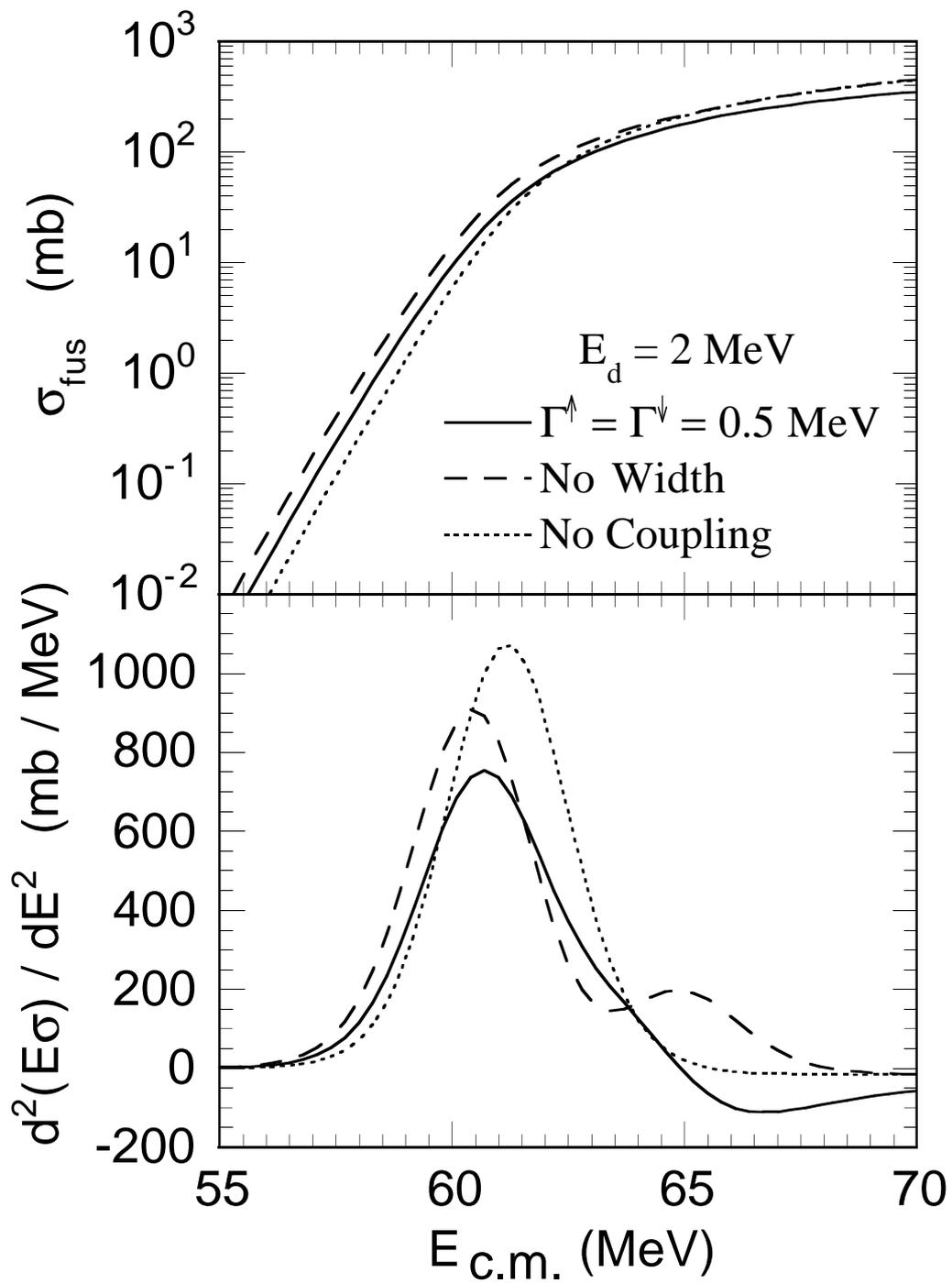
\end{document}